\documentclass[aip,superscriptsize,reprint,onecolumn]{revtex4-1}
\usepackage[utf8]{inputenc} 

\usepackage{graphicx,xcolor,overpic,mathtools}
\usepackage{amsthm,amsmath,amssymb}
\usepackage[colorlinks]{hyperref}
\definecolor{coolblack}{rgb}{0.0, 0.18, 0.39}
\hypersetup{
    colorlinks = true,
    citecolor = {coolblack},  
    urlcolor = {blue}
   }
\PassOptionsToPackage{normalem}{ulem}
\usepackage{ulem} 
\usepackage{sidenotes}
\usepackage{bm}
\usepackage{url}
\usepackage{physics}
\usepackage[makeroom]{cancel}
\usepackage{tensor}
\usepackage{mathrsfs}
\usepackage{slashed}
\usepackage{tikz}
\usepackage[mode=buildnew]{standalone}

\newcommand{\comment}[1]{}

\NewDocumentCommand{\evat}{sO{\bigg}mm}{%
  \IfBooleanTF{#1}
   {\mleft. #3 \mright|_{#4}}
   {#3#2|_{#4}}%
}
\usepackage{enumerate}
\usepackage{cleveref}

\def\shrinkage{0mu}
\def\vecsign{\mathchar"017E}
\def\dvecsign{\smash{\stackon[-2.1pt]{\mkern-\shrinkage\vecsign}{\rotatebox{180}{$\mkern-\shrinkage\vecsign$}}}}
\def\dvec#1{\def\useanchorwidth{T}\stackon[-4.2pt]{#1}{\,\dvecsign}}
\usepackage{stackengine}
\stackMath

\begin{document}
\title[
]{Constrained instanton approximation of Skyrmions with massive pions}

\author{Alberto García Martín-Caro}
\email{alberto.martin-caro@usc.es}

\affiliation{Departamento de F\'isica de Part\'iculas, Universidad de Santiago de Compostela and Instituto
Galego de F\'isica de Altas Enerxias (IGFAE) E-15782 Santiago de Compostela, Spain}

\date[ Date: ]{\today}
\begin{abstract}
We present the idea of using the holonomy along a line of a constrained instanton solution (an approximate solution of the Euclidean equations of motion subject to a constraint) of an $\rm SU(2)$ Yang-Mills-Higgs theory to approximate the Skyrmion solution in a chiral model with massive pions.
The fact that the gauge field acquires a nonzero mass due to the Higgs mechanism implies that a constrained instanton decays exponentially far from its center, and so does the Skyrmion configuration that it generates via the Atiyah-Manton construction. This is precisely the desired behavior at large distances for the true Skyrmion solutions when the pion mass is included.
\end{abstract}
\maketitle
\comment{
\begin{minipage}{\textwidth}
\tableofcontents
\end{minipage}
}
\section{Introduction}
The vacuum structure of an $\rm SU(2)$ Yang-Mills theory in $3+1$ dimensions is nontrivial, in the sense that it presents a discrete set of degenerate classical minima. These minima are associated to static (in the temporal gauge), pure gauge configurations of the form $A_\mu=\Omega^{-1}(\mathbf{x})\partial_\mu\Omega(\mathbf{x})$, with ${\rm SU(2)}\ni\Omega(\mathbf{x})\rightarrow1$ as $|\mathbf{x}|\rightarrow \infty$. Mathematically, these configurations can be seen as continuous mappings $\mathbb{R}^3\cup \{\infty\}\sim S^3\rightarrow {\rm SU(2)}\sim S^3$, which are classified by their third homotopy group, $\pi_3(S^3)=\mathbb{Z}$. In other words, vacuum gauge field configurations are separated in different homotopy classes, labeled by an integer $n[\Omega]$, which can be obtained as an integral
\begin{equation}
    n[\Omega]=-\frac{1}{24 \pi^{2}} \int_{S^3} d^{3} x \epsilon^{i j k} \operatorname{Tr}\left(\Omega ^{-1}\partial_{i} \Omega \Omega^{-1} \partial_{j} \Omega \Omega^{-1} \partial_{k} \Omega\right).
\end{equation}

Vacuum configurations with different topological degree cannot be continuously deformed into each other without generating non-vacuum gauge fields, so these vacua are separated by a potential barrier in the quantum theory. The tunneling transition between classically degenerate vacua of a quantum theory is generally described by instantons, which are solutions of the Euclidean field equations, i.e absolute minimizers of the Euclidean action
\begin{equation}
    S_{\text E}=\frac{1}{4g^2}\int \Tr(F^a_{\mu\nu} F^a_{\mu\nu})d^4x,
\end{equation}
with $F^a_{\mu\nu}=\partial_\mu A^a_\nu-\partial_\nu A^a_\mu +\epsilon^{abc}A^b_\mu A^c_\nu$ is the gauge field strength tensor, and $A_\mu=-ig\tfrac{\tau^a}{2}A^a_\mu$ is the Lie algebra-valued Yang-Mills field.

Instanton solutions are localised in spacetime, and can also be classified by their homotopy class when seen as mappings on $\mathbb{R}^4$ that asymptote to pure gauge configurations on its boundary, $S^3_\infty$ (so that their action remains finite). Therefore, we can also think of Yang-Mills instantons as topological solitons in the Euclidean version of the theory, with topological charge (instanton number) given by
\begin{equation}
    n=-\frac{1}{16\pi^2}\int d^4x\Tr\,( F_{\mu\nu}\Tilde{F}_{\mu\nu}).
\end{equation}

These tunnelling solutions are usually necessary in order to understand various non-perturbative processes in the Standard Model and other (super)Yang-Mills theories. In particular, they play an important role in Quantum Chromodynamics (QCD), where they have been used to solve the axial $\rm U(1)$ problem \cite{tHooft:U1} or to explain chiral symmetry breaking \cite{Diakonov:1995ea}.  

Further, $\rm SU(2)$ Yang-Mills instantons are also related to solitons in the Skyrme model, a relativistic, $3+1$ dimensional chiral effective field theory first proposed by Tony Skyrme \cite{Skyrme:1961vq}. The Skyrme model describes the dynamics of self-interacting pion fields, which are collected into an $\rm SU(2)$-valued field $U(t,\mathbf{x})$. Baryons are described in this model as topological solitons, called Skyrmions, whose topological charge is identified with baryon number. In a seminal paper \cite{ATIYAH1989438}, M. Atiyah and N. Manton proposed that a (static) Skyrme field configuration $U(\vec{x})$ with baryon number $k$ may be obtained as the holonomy of an $\rm SU(2)$ Yang-Mills $k$-instanton configuration along the Euclidean time direction, formally
\begin{equation}
    U(\mathbf{x})=\mathcal{P}\exp(-\int^\infty_{-\infty}A_4(\mathbf{x},x_4)dx_4),
    \label{Holonomy}
\end{equation}
with $\mathcal{P}$ denoting path ordering of the exponential. Such a construction is almost gauge invariant, since the effect of a gauge transformation of $A_\mu$ by $g(x)$ is just a conjugation of $U$ by the asymptotic value $g(\infty)$. This can be understood as an isospin
rotation of the corresponding Skyrme field.

Moreover, the 
moduli space of $\rm SU(2)$ instanton configurations of charge  $k$, including the choice of gauge at infinity, is a connected manifold of 
dimension $8k$ \cite{AtiyahHitchin,JACKIW1977189}. Hence, the set of Skyrme fields with $B=k$ obtained via the Atiyah-Manton construction will be an $8k-1$ dimensional manifold, as a global time translation of the instanton leaves  the Skyrme field unchanged. This manifold is sometimes called the \emph{moduli space of $B=k$ instanton-generated Skyrmions}, since even though the Skyrme fields so-constructed are not exact solutions to the Skyrme field equations, some are good approximations to minimal energy Skyrmions and other
field configurations describing their low energy dynamics in that topological sector. The Atiyah-Manton approximation was further developed in \cite{Leese:1993mc, Hosaka:1991kh,Sutcliffe:1994wx}, and it was also extended in order to include instantons at finite temperature, \cite{Cork:2018sem,Cork:2021uov}
and skyrmions from gravitational instantons  \cite{Dunajski:2012fg}.
Its application to describe low energy interactions between nuclei was studied shortly after it was first proposed \cite{Atiyah:1992if,Hosaka:1991kh}, and has been put forward recently in 
\cite{Halcrow:2021wwc,Halcrow:2022bxw}.

The useful idea that Skyrmions can be seen as instantons in a higher dimensional space has been thoroughly studied in the holographic context. In Sutcliffe \cite{Sutcliffe:2010et}, it is explicitly shown that the Skyrme model appears, together with a Kaluza-Klein tower of vector mesons, after a dimensional reduction of a $4+1$ dimensional Yang-Mills theory. This phenomenon is also realized in several holographic QCD models. In particular, the Sakai-Sugimoto model, a holographic dual of QCD with $N_f$ massless quarks proposed in \cite{Sakai:2004cn}. The low energy effective action of such a model consists of a five-dimensional Yang-Mills and Chern-Simons theory on a curved background. In this model, the massless pion and an infinite tower of massive (axial-)vector mesons are interpreted as Kaluza-Klein states associated with the  holographic direction, and baryons are identified as D4-branes wrapped on a non-trivial four-cycle in the D4 background. Such a D4-brane is realized as an instanton configuration in the worldvolume gauge theory. Also, the pion effective action obtained from the KK reduction of the $4+1d$ Yang-Mills theory is precisely that of the Skyrme model, in which baryons appear as solitons. It can be shown that the baryon number of a Skyrmion, i.e the winding number carried by the pion field, is equivalent to the instanton number in the Yang-Mills theory. In this way, the Sakai-Sugimoto model presents two dual descriptions of baryons, as wrapped D-branes (instantons) and as Skyrmions in the KK-reduced theory.

Apart from QCD-like theories,  instanton-mediated processes are also important in more complicated gauge theories presenting the Higgs mechanism, such as the electroweak sector of the Standard Model, where they are associated to the violation of baryon number \cite{Espinosa:1989qn}. Such transitions become relevant only at very high energies and have been proposed as a plausible source of baryon number violation leading to baryogenesis in the early universe \cite{Rubakov:1996vz}. However, the Higgs mechanism poses a slight complication concerning instantons in these gauge theories. Indeed, there are no solutions to the Euclidean field equations, i.e. no exact minima of the Euclidean action in sectors with nonzero topological degree. The reason, as we will discuss below, is that the action for instanton-like configurations, with a nontrivial Higgs field, depends on the instanton size and decreases as it tends to zero. To evaluate the functional integral in that case one introduces a constraint that fixes the size of the configuration \cite{tHooft76}, then minimizes the action under this constraint and finally integrates over all possible values of the instanton size. The constrained solutions that one uses to calculate instantonic contribution to the path integral in the Higgs phase are called \emph{constrained instantons}.

\section{The Skyrme model and the Atiyah-Manton construction}
The Lagrangian of the Skyrme model is given by
\begin{equation}
    \mathscr{L}=-\frac{f_\pi}{16}\Tr(L_\mu L^\mu)+\frac{1}{32e^2}\Tr([L_\mu,L_\nu]^2)+\frac{1}{8}m_\pi^2f_\pi^2\Tr(U-\bm{1}),
    \label{Sklag}
\end{equation}
where $L_\mu=U^\dagger\partial_\mu U$ is the associated left-invariant Maurer-Cartan form. The lagrangian depends on the three parameters $f_\pi,m_\pi$ and $e$, namely, the pion decay
constant, the pion mass and the (dimensionless) Skyrme parameter. The first two constants are common to every chiral effective theory involving pions, and can be fixed by low energy constraints on their properties. The latter is associated to the fourth-order Skyrme term, which allows for stable soliton solutions.
Such solutions are called Skyrmions, and correspond to field configurations that minimize the static energy functional
\begin{equation}
    E=-\int \Big[\frac{1}{2}\Tr(L_i L_i)+\frac{1}{16}\Tr([L_i,L_j]^2)+m^2\Tr(U-\bm{1})\Big]d^3x,
    \label{SkEn}
\end{equation}
where we have changed our energy and lenght units to $f_\pi/4e$ and $2/ef_\pi$, so that the dimensionless pion mass parameter is related to the pion mass $m_\pi$ by $m=2m_\pi/ef_\pi$.

Furthermore, Skyrmions are topological solitons, since any finite energy configuration must satisfy the boundary condition ${U(\mathbf{x})\rightarrow \bm{1}}$ as $|\mathbf{x}|\rightarrow \infty$. Thus, static skyrmions are maps $U:\mathbb{R}^3_\infty\sim S^3\mapsto {\rm SU(2)}\sim S^3$, classified by their third homotopy group, $\pi_3(S^3)= \mathbb{Z}$, into homotopy classes labelled by their corresponding topological degree, which can be written in terms of $L_\mu$ as
\begin{equation}
    B=\int B_0 d^3 x\in \mathbb{Z},\qquad B_\mu=\frac{1}{24\pi^2}\varepsilon_{\mu\nu\rho\sigma}\Tr(L^\nu L^\rho L^\sigma).
\end{equation}
In this work, we will be primarily interested in the instanton-generated $B=1$ Skyrmion.
Indeed, the $B=1$ Skyrmion solution is well known, and can be written using the so-called hedgehog configuration
\begin{equation}
    U(\mathbf{x})=e^{if(\rho) \mathbf{n}^a\sigma_a},\qquad\text{where} \qquad \rho=|\mathbf{x}|, \quad \vb{n}=\frac{\mathbf{x}}{\rho},
\end{equation}
and $f(\rho)$ is the radial profile of the Skyrmion, which, for the $B=1$ case, must satisfy $f(0)=\pi$ and $f(\infty)=0$. Introducing the hedgehog ansatz into the static energy functional, one obtains
\begin{equation}
    E=4\pi\int\left(\rho^2 f^{\prime 2}+2\left(f^{\prime 2}+1\right) \sin ^2 f+\frac{\sin ^4 f}{\rho^2}+2 m^2 \rho^2(1-\cos f)\right) d \rho,
\end{equation}

On the other hand, the (Euclidean, $4d$) pure Yang-Mills action presents instanton solutions which, in the 't Hooft gauge, read
\begin{equation}
    A_\mu(x)=\frac{1}{2}\sigma_{\mu\nu}\partial_\nu \log\alpha(x),
    \label{pureinst}
\end{equation}
with $\alpha$ a solution of the Laplace equation in $\mathbb{R}^4$, $\partial^2 \alpha=0$, and $\sigma_{\mu\nu}$ the $\mathfrak{so(4)}$ Lie algebra generators, defined as \footnote{we are using the conventions in \cite{Espinosa:1989qn}} 
\begin{equation}
    \sigma_{\mu\nu}=\frac{1}{4i}(\sigma_\mu\sigma_\nu^\dagger-\sigma_\mu^\dagger\sigma_\nu),\qquad \text{with}\quad \sigma_\mu=(\bm{\tau},i\bm{1}_2).
\end{equation}
In particular, the $k=1$-instanton is given by 
\begin{equation}
    \alpha(x)=1+\frac{\lambda^2}{|x-X|^2}= 1+t,\quad t\doteq\frac{\lambda^2}{|x-X|^2},
\end{equation}
with $\lambda\in \mathbb{R}$, $X\in \mathbb{R}^4$ the moduli parametrizing the size and position of the instanton in Euclidean space, respectively.
These $1$-instantons generate, after calculating the corresponding holonomy as in \eqref{Holonomy}, a seven-dimensional manifold of B = 1 Skyrme hedgehogs, the coordinates being the position, orientation and scale size, and whose radial profile is given by 
\begin{equation}
    f(\rho)=\pi\left[1-\left(1+\frac{\lambda^2}{\rho^2}\right)^{-1 / 2}\right].
\end{equation}
Although instantons are scale invariant, the energy of the associated Skyrmion configurations depends on the particular value of the (in principle, arbitrary) parameter $\lambda$. Therefore, its value must be fixed so that the energy of the resulting Skyrme field is minimized, which reduces the moduli space by one dimension. It turns out that, for the massless case, the value of this scale is $\lambda^2 = 2.11$, with an energy of $E^{\rm inst}_{m=0} = 1.243\times 12\pi^2$, which is only $1\%$ above that of the true Skyrmion solution.

\section{Constrained instantons in {Yang-Mills-Higgs} theories}
We now consider an $\rm SU(2)$ Yang-Mills-Higgs theory on four dimensional Euclidean space $\mathbb{R}^4$, given by the Lagrangian
\begin{equation}
    \mathscr{L}=\frac{1}{g^2}\qty{\frac{1}{4}F_{\mu\nu}^aF_{\mu\nu}^a+\kappa \qty[(D_\mu \phi)^\dagger D_\mu\phi +\frac 1 4 (\phi^\dagger\phi-\mu^2)  ]},
    \label{lagYMH}
\end{equation}
where $D_\mu=\partial_\mu -i \frac{\tau ^a}{2}A^a_\mu\equiv \partial_\mu +A_\mu$ is the associated covariant derivative, $\phi$ an $\rm SU(2)$ Higgs doublet $g$ and $\kappa$ the gauge and Higgs couplings, respectively. Finally, $\mu^2$ is the vacuum expectation value of the Higgs field in the spontaneously broken phase, in which the gauge field acquires a mass $m=\sqrt{\tfrac{\kappa\mu^2}{2}}$ due to the Higgs mechanism.

For $\mu=0$, the Yang-Mills-Higgs field equations present instanton solutions which are the same as in the pure Yang-Mills case, with $\phi = 0$.  On the other hand, for $\mu\neq 0$, the nontrivial vacuum expectation value of the Higgs field breaks conformal symmetry, and there cannot exist an exact instanton solution of the coupled field equations. This can be easily seen by applying Derrick's scaling argument \cite{AFFLECK1981429}. Indeed, rescaling any instanton field configuration with $\phi\neq 0$ as
\begin{equation}
    A_\mu(x)\rightarrow aA_\mu(ax),\quad \phi(x)\rightarrow \phi(ax),
\end{equation}
the action becomes $(y=ax)$:
\begin{equation}
    \hspace{-0.5cm}
    S=\int d^4y\frac{1}{g^2}\qty{\frac{1}{4}F_{\mu\nu}^a(y)F_{\mu\nu}^a(y)+\frac{\kappa}{a^2} \qty[(D_\mu \phi(y))^\dagger D_\mu\phi(y) +\frac{1}{4a^2} (\phi^\dagger\phi-\mu^2)  ]},
\end{equation}
and since all terms in the integrand are positive, the action can always be made smaller by taking the limit $a\rightarrow \infty$. Since the expression for $aA_\mu( ax)$ is exactly equal to $A_\mu(x)$ with $\lambda$ replaced by $\lambda/a$, we get that the action-minimizing instanton has zero size. However, the nontrivial topology of the space of field configurations is not affected by the Higgs mechanism, so that instanton-like field configurations with nonzero topological degree still exist as approximate solutions of the field equations. These are the so-called \emph{constrained instantons}, and have been used to approximately calculate nonperturbative contributions to the path integral in the standard model and (super) Yang-Mills theories at the Higgs phase \cite{tHooft76, Nielsen:2005mh, Rubakov:1996vz}. 

The constrained instanton idea is based on the existence of a finite number of destabilizing directions in the gauge field space, along which the action always decreases, so that one introduces constraints that prevent deformations in these directions. Then, one just minimizes the action for fields subject to such constraints. For the spontaneously broken phase of $\rm SU(2)$ Yang-Mills-Higgs theory, the destabilizing direction is that parametrized by the rescaling parameter $a$ (note that this would correspond to a zero mode in the pure Yang-Mills phase), so one needs to impose a constraint that fixes the instanton size $\lambda$ at some value.\footnote{Also, the constraint must allow finiteness of the action, which is a nontrivial demand for choosing a valid constraint, see \cite{Nielsen:1999vq}.} The contribution to the path integral coming from these configurations is then obtained by integration over all (fixed) values of $\lambda$. Remarkably, the Higgs mechanism resolves the infrared divergence problem due to large instantons dominating the path integral measure on an otherwise scale-symmetric field theory such as pure Yang-Mills, since these become suppressed on the Higgs phase \cite{tHooft76, Espinosa:1989qn,shifman_2022}.

The Yang-Mills-Higgs equations coming from the Lagrangian \eqref{lagYMH} are
\begin{align}
    D_\mu F_{\mu\nu}^a+\frac{\kappa}{2}\qty(-i \phi^\dagger\tau^a \dvec{\partial}_\mu\phi-A^a_\mu \phi^\dagger \phi)&=0,\label{YMHeq1}\\
    D^2\phi-\phi(\phi^\dagger \phi-\mu^2)&=0.
    \label{YMHeq2}
\end{align}

One could naively try to find instantonic solutions to the Yang-Mills-Higgs equations. The finite action condition in the broken phase implies that gauge fields must become pure gauge configurations asymptotically, and the Higgs field must tend to a point in the corresponding vacuum manifold, 
\begin{equation}
    A_\mu(x)\rightarrow \Omega\partial_\mu\Omega^{-1}, \quad \phi(x)\rightarrow \Omega\phi_{\rm vac},\quad x\rightarrow \infty,\quad \Omega(x)\in {\rm SU(2)}
\end{equation}
with $\phi_{\rm vac}$ a general base point of the Higgs potential vacuum manifold. Without loss of generality, we may choose $\phi_{\rm vac}=(0,\mu)^\intercal$.
Therefore, a general ansatz for the (spherically symmetric) $k=1$ instanton would be of the form
\begin{equation}
    A_\mu= \tilde\xi(x)\Omega \partial_\mu \Omega^{-1},\qquad \phi(x)=(1-\chi(x))\Omega\phi_{\rm vac},
    \label{BPSTconstrainedansatz}
\end{equation}
where $\Omega=i\frac{\sigma_\mu x^\mu}{\sqrt{x_\mu x_\mu}}$ is an element of the first nontrivial homotopy class in $\pi_3(\rm SU(2))$, and $\tilde{\xi},\chi$ are functions of the radial euclidean coordinate $r=|x|$ which satisfy the conditions ${\tilde{\xi},\chi\rightarrow 1,0}$ as $r\rightarrow \infty$. This ansatz corresponds to the generalization of the BPST instanton solution \cite{Belavin:1975fg} in the regular gauge to the Higgs phase, and was used in \cite{Klinkhamer:1993kn} to find a constrained instanton solution using a specific constraint. 
In particular, for $\mu=0$, i.e. the pure Yang-Mills case, the instanton solution is given in this gauge by $\tilde \xi=r^2/(r^2+\lambda^2)$.

However, we will find it useful to gauge-transform \eqref{BPSTconstrainedansatz} into a new form, in which $\phi\rightarrow\phi_{\rm vac}$. This can be achieved with a gauge transformation by $\Omega^{-1}$, which brings the ansatz into the so-called singular gauge:
\begin{equation}
    A_\mu(x)=\frac{x^\nu\sigma_{\mu\nu}}{x^2}\xi(r)= -\sigma_{\mu\nu}\partial_\nu a(r),\quad 
    \phi(x)=(1-\chi(x))\phi_{\rm vac},
        \label{singulargaugeansatz}
\end{equation}

where $\xi$ is a function related to the original $\tilde{\xi}$, and ${\xi=a'/r}$.  With this gauge choice, the boundary conditions 
$
    {A_\mu(x)\xrightarrow{x\rightarrow\infty} 0, \, \phi(x)\xrightarrow{x\rightarrow\infty} (0,\mu)^\intercal, }
$
are satisfied, and
 \cref{YMHeq1,YMHeq2} linearize for $A_\mu(x)$ and $\chi(x)$ in the $x\rightarrow\infty$ limit,
\begin{align}
    \qty[-\delta_{\mu\nu}\partial^2+\partial_\mu\partial_\nu+m^2]A_\mu&=0,\qquad m=\sqrt{\frac{\kappa\mu^2}{2}}
    \\
    (-\partial^2+\mu^2)\chi&=0.
\end{align}
Any solution of the linearized equations that satisfies the correct boundary conditions is given by
\begin{equation}
    \chi \propto G(x;\mu),\qquad  A_\mu \propto \sigma_{\mu\nu}\partial_\nu G(x;m),
\end{equation}
with 
\begin{equation}
    G(x;\alpha)=\frac{\alpha}{r}K_1(\alpha r)
    \label{GreenBessel}
\end{equation}
the Green function associated to the four dimensional, Euclidean Klein-Gordon operator, $-\partial^2+\alpha^2$, and $K_n(z)$ is the modified Bessel function of degree $n$.
\subsection{Perturbative construction}
Unfortunately, as previously shown using Derrick's argument, the boundary conditions we have imposed are not compatible with a regular solution of the full system of Yang-Mills-Higgs equations. Nonetheless, it was argued in \cite{AFFLECK1981429} that a 
perturbative  solution  can  be  constructed  if we  add  extra  terms  to  the 
right-hand  sides  of  such equations, with  coefficients that  can  be  adjusted 
order by order in perturbation  theory to obtain the desired boundary conditions. The presence of these extra terms is equivalent to adding a constraint at the level of the action, so the obtained solutions are \emph{constrained instantons}. This procedure was further extended in \cite{Nielsen:1999vq} with the additional requirement that the perturbative solution yields a finite value for the action functional, and a constrained instanton solution in the Higgs phase of $\rm SU(2)$ YMH theory was calculated perturbatively in the parameter $\mu$ up to $\order{\mu^4}$. The solution looks like the standard self-dual instanton near its core, but decays exponentially (instead of polynomially) as $\exp(-\mu r)$ far from the center. We will now reproduce the construction of such solution, following the same perturbative method.

We start with the singular gauge ansatz \eqref{singulargaugeansatz}, and define two real functions $\alpha=\exp(a)$ and $f=\mu(1-\chi)$ depending on $x$ in terms of the parameter $t=\tfrac{\lambda^2}{r^2}$. We also expand these functions as
\begin{equation}
    \alpha=\sum\limits_{n=0}^\infty \alpha_{2n},\qquad f=\sum\limits_{n=0}^\infty f_{2n+1},
\end{equation}
where the subscript on each function indicates the lowest power of $\mu$ appearing on it. Introducing this ansatz into the Yang-Mills-Higgs Lagrangian, the equations of motion for $\alpha$ and $f$ are ($'=d/dt$) \cite{Nielsen:1999vq}:
\begin{equation}
    (\alpha^{-3}t^3\alpha'')'=\frac{\kappa\lambda^2}{8}f^2\alpha^{-3}\alpha',\quad \alpha^2f''-\frac{3}{4}(\alpha')^2f=\frac{\alpha^2\lambda^2 f}{8t^3}(f^2-\mu^2),
\end{equation}
\begin{figure}
    \centering
    \includegraphics[scale=0.7]{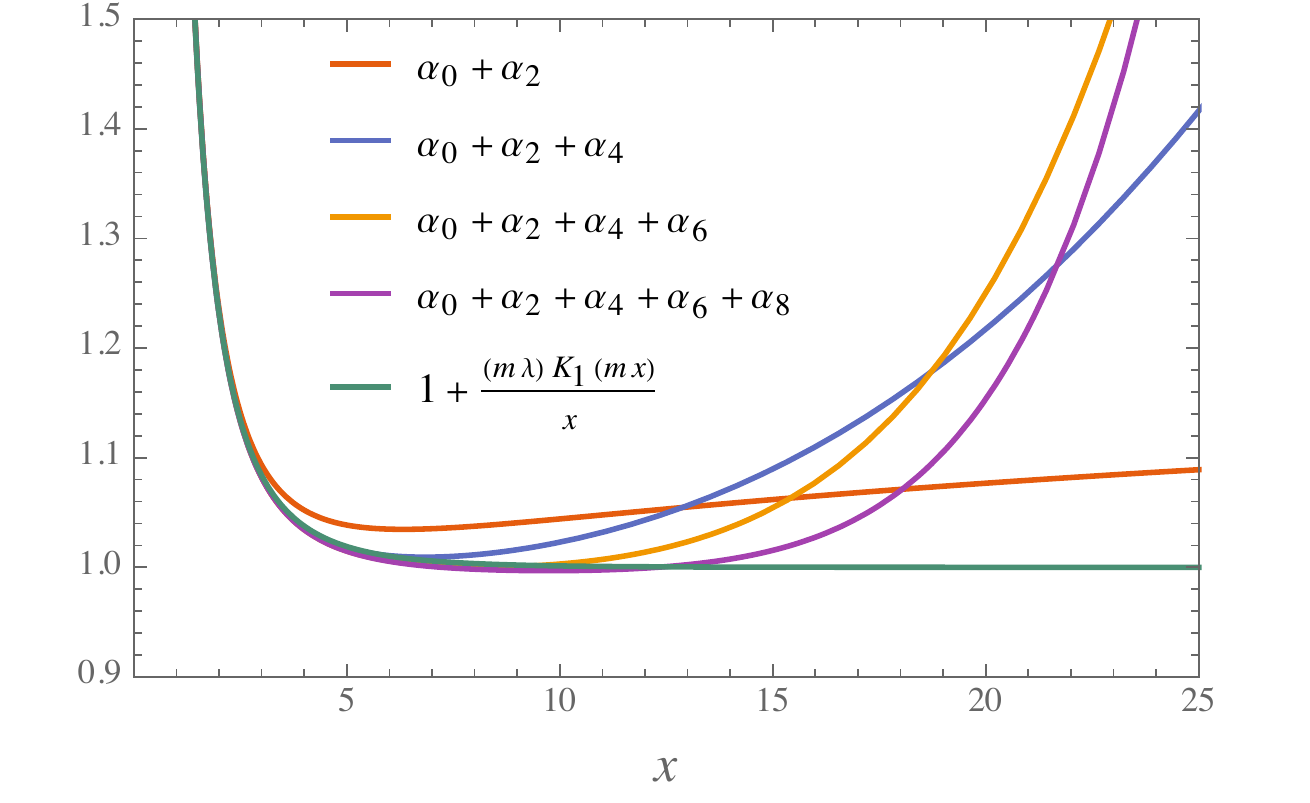}
    \caption{Perturbative solution of the constrained instanton profile at different orders in the Higgs vev parameter, for $\mu^2=m^2=0.1$. The large $x$ limit towards it needs to converge is also plotted in green. }
    \label{fig:alphas}
\end{figure}
which can be solved by quadrature at each order. The required boundary conditions plus finiteness of the action require that these equations are modified order by order by adding certain terms at the rhs. These terms (and the corresponding constraint) are computed at each order in $\mu$, and are unique up to $\order{\mu^4}$ \cite{Nielsen:1999vq}. This can be done iteratively, starting with the zeroth-order instanton solution, $\alpha_0(t)=1+t$, and substituting in the equation for $f$, which yields $f_1=\mu/\sqrt{1+t}$. This solution already satisfies the boundary condition for $\phi$, so no modification of the equation for $f$ is needed at this order. In fact, the equation for $f$ does not need to be modified at any order \cite{Nielsen:1999vq}. On the other hand, at second order, the modification of the equation for $\alpha$ turns out to be unique, (i.e. the corresponding constraint is unique up to $\order{\mu^4}$), and therefore, one can solve for $\alpha_2$ unambiguously,
\begin{equation}
    \alpha_2(t)=\frac{\lambda^2 m^2}{4}\qty[\log{\frac{\lambda^2 m^2}{4t}}+2\gamma-1],
\end{equation}
where $\gamma$ is Euler's constant, appearing in the small $m$ expansion of the modified Bessel function in \eqref{GreenBessel}. In order to find the next order term in $\alpha$, one needs to solve first for the Higgs field at order three. As we are not interested in its explicit form, let us skip it and just give the result for the next order term in $\alpha$, i.e. the $\order{\mu^4}$ term. It's found in \cite{Nielsen:1999vq} to be
{
\begin{align}
    &\alpha_4(t)=\frac{\lambda^4 m^4}{64 t}\Bigg\{\left(-2 t^2+4 \gamma -5\right)+2 \Bigg[t \log (t) \Big(  \frac{\mu^2}{m^2}(4 \gamma+3)+\notag\\&2 \left(\frac{\mu^2}{m^2}-1\right) \log \left(\frac{\lambda^2 \mu^2}{4}\right)-2 \log \left(\frac{m^2}{\mu^2}\right)-4 \gamma  +1\Big)+\notag\\
    +&t (t+1) \log \left(\frac{1}{t}+1\right) \left(4 \log \left(\frac{\lambda^2
   m^2}{4 t}\right)+(t+8 \gamma -9)-\frac{\mu^2}{m^2} (t-5)\right)+\notag\\
   +&\log\left(\frac{\lambda^2 m^2}{4 t}\right)\Bigg]-4 t \Phi \left(\frac{1}{t}\right) \left(\frac{\mu^2}{m^2}(1-2t)+4t+1+2 \frac{\mu^2}{m^2} t^2\right)\Bigg\},
\end{align}}
with 
\begin{equation}
    \Phi(x)=\int_0^x\frac{\log (1+u)}{u}du.
\end{equation}
For simplicity, in the following we will make the Higgs field vev equal to the mass (i.e. we fix the Higgs coupling constant $\kappa=2$
). Doing so, the expression above simplifies to 
\begin{align}
    \alpha_4(t)&=\frac{\lambda^2 m^2}{32 t} \Bigg\{ \log \left(\frac{\lambda^2m^2}{4 t}\right)-4 t (t+1) \Phi \left(\frac{1}{t}\right)+2 \gamma -\frac 5 2+\notag\\+&4 t \left[(t+1) \log \left(\frac{1}{t}+1\right) \left(\log \left(\frac{\lambda^2m^2}{4 t}\right)+2 \gamma -1\right)+\log (t)\right]\Bigg\}.
\end{align}
Furthermore, the authors in \cite{Nielsen:1999vq} found a recurrence relation for obtaining the leading term on each higher order component $\alpha_{2n}$
\begin{equation}
    \frac{d}{d t} t^{3} \frac{d^{2} \alpha_{n}}{d t^{2}} \simeq \frac{\lambda^{2} m^2}{4} \frac{d \alpha_{n-2}}{d t},
    \label{alpha_recurrence}
\end{equation}
whose solutions coincide, at each $n$, with the $n$-th order term in the small $m$ expansion of $G(x;m)$. The convergence towards $G(x,m)$ of $\alpha(t)$ at large distances for different orders in $\mu^2$ is shown in \eqref{fig:alphas}. The $\alpha_6$ and $\alpha_8$ are taken as the corresponding solutions to \eqref{alpha_recurrence}.
The perturbative solution is unsatisfactory for our purposes, because it will always diverge for some value of $r$ at any given truncation order (as we can see in \cref{fig:alphas}). After computing the corresponding holonomy, this translates into a Skyrme field that does not satisfy the correct boundary conditions at infinity. A simple  way around this problem is just to consider the asymptotic solution, namely

\begin{equation}
    \alpha(x)=1+\frac{\lambda^2m}{r}K_1(mr),
\label{alphafin}
\end{equation}
which does of course yield well-behaved Skyrmions at infinity, as we will show in the next section.

\subsection{Non-perturbative approach}

Apart from the perturbative method, an alternative approach for constructing constrained instanton configurations in Yang-Mills-Higgs theory was used in \cite{Wang:1994rz}, in which instead of fixing the constraint and trying to solve the equation for the corresponding constrained instanton, the reverse method is followed. One chooses a particular functional form for the constrained instanton, and the corresponding constraint may be systematically calculated afterwards due to the freedom in choosing the constraint in the first place.
The only requirement in choosing the constrained instanton shape a priori is that it satisfies some constraint-independent
boundary conditions at the origin and far from its center.
Indeed, if one considers again the BPST-like ansatz, the (dimensionless) functions $\xi,\chi$ can depend on $r=|x|$ only through two dimensionless combinations, $\lambda/r$ and $m r$. We may expand these functions both for small $(r\ll \lambda)$ and large $(r\gg m^{-1})$ distances in terms of these two combinations, as \cite{Espinosa:1989qn,Wang:1994rz}:
\begin{align}
    \xi(r)&=\xi_0(\tfrac{\lambda}{r})+(m r)^2 \xi_1(\tfrac{\lambda}{r})+\cdots =\notag\\
    &=(\tfrac{\lambda}{r})^2\xi^0(m r)+ (\tfrac{\lambda}{r})^4\xi^1(m r)+\cdots \label{expansion1},\\\notag\\
    \chi(r)&=\chi_0(\tfrac{\lambda}{r})+(m r)^2\ln(m r) \chi_1(\tfrac{\lambda}{r})+\cdots =\notag\\
    &=(\tfrac{\lambda}{r})^2\chi^0(m r)+ (\tfrac{\lambda}{r})^4\chi^1(m r)+\cdots.
    \label{expansion2}
\end{align}
Thus for recovering the pure Yang-Mills solution in the $m\rightarrow 0$ limit we require 
\begin{equation}
    \xi_0(\tfrac{\lambda}{r})=\frac{2\lambda^2}{r^2+\lambda^2},\qquad \chi_0(\tfrac{\lambda}{r})=1-\sqrt{\frac{r^2}{r^2+\lambda^2}}\label{limit1},
\end{equation}
where $\chi_0$ above corresponds to the scalar field profile in the instanton background without backreaction \cite{tHooft76}.

On the other hand, the linearized equations at large distances impose
\begin{equation}
        \xi^0(mr)=m^2r^2K_2(mr),\qquad \chi^0(mr)=\tfrac{1}{2}m rK_1(mr).\label{limit2}
\end{equation}

A pair of functions proposed by Wang in \cite{Wang:1994rz} satisfying \cref{expansion1,expansion2,limit1,limit2} are 

\begin{equation}
\xi(r)=\frac{ \lambda^2 m^2 K_2\left(mr\right)}{[1+\lambda^2 m^2 K_2\left(mr)/2\right] },\quad \chi(r)=\qty[1-\left(\frac{r^2}{r^2+\lambda^2 m r K_1\left(mr\right)}\right)^{\frac{1}{2}}].
\label{Wang}
\end{equation}
Obviously, the expressions in \cref{Wang} are associated with a particular constraint at the level of the action, which could be calculated, at least formally. Of course, they are not unique, and a different pair of functions satisfying the correct boundary conditions could have been chosen. 

\section{Skyrmions from constrained instanton configurations}
Despite its success in describing the moduli space of massless pion Skyrmions, the instanton approximation does not work in the massive pion case. This can be understood by noting that the asymptotic decay of the instanton generated profile function is $f \sim \lambda ^2/(2\rho^2)$, which has the correct form for massless pions, but decays too slowly for the massive pion case, ultimately generating configurations with infinite energy. Thus this form of the instanton holonomy method is not applicable to massive pions.

However, as we have seen in the previous section, we are still able to find instanton-like configurations in Yang-Mills-Higgs theory that preserve their topological properties and nevertheless present an exponential radial decay due to the Higgs mechanism. Hence, constrained instantons are natural candidates for generating exponentially decaying Skyrme hedgehogs via the Atiyah-Manton construction.  
Indeed, from the ansatz in the previous section we get
\begin{equation}
    A_4(\mathbf{x},x_4)=-in^a\tau_ a\qty(\rho\frac{\lambda^2}{(x_4^2+\rho^2)^2}\frac{\alpha'}{\alpha}),
\end{equation}
with $\alpha(t)$ given by \eqref{alphafin}. The profile function generated by the constrained instanton approximation can therefore be written 
\begin{equation}
    f(\rho)=\int_0^\infty\frac{\rho\lambda^2}{(x_4^2+\rho^2)^2}\frac{\alpha'(t)}{\alpha(t)}dx_4.
\end{equation}
On the other hand, if we choose to use the nonperturbative configuration generated by the functions \eqref{Wang} for constructing the constrained instanton, the profile function of the corresponding Skyrmion can be directly obtained as 
\begin{equation}
        f(\rho)=\int_0^\infty\frac{\rho\lambda^2m^2K_2(m\sqrt{x_4^2+\rho^2})}{2(x_4^2+\rho^2)[1+\lambda^2 m^2K_2(m\sqrt{x_4^2+\rho^2})/2]}dx_4.
\end{equation}

As pointed out in \cite{Espinosa:1989qn}, the Lagrangian \cref{lagYMH} is invariant under independent global rotations of $A_\mu$ and $\phi$, even after substituting the ansatz \eqref{singulargaugeansatz}.
Thus, in the constrained instanton the gauge and Higgs
field orientations are not correlated. Furthermore, the Higgs field orientation is completely fixed
once we choose a specific point in the vacuum manifold, $\phi_{\rm}$ as the vacuum state in the spontaneously broken phase. On the other hand, the orientation of $A_\mu$ remains a symmetry of the Yang-Mills-Higgs action and therefore has to be treated as a collective coordinate along with the position of the constrained instanton. Of course, these collective coordinates will be inherited by the Skyrmion field configurations generated by constrained instanton configurations, so that the structure of the instanton-generated Skyrmion moduli space is preserved even after the addition of the Higgs field.

Obviously, the constrained instanton solution has lost its scale invariance due to a finite Higgs vev, so the scale parameter $\lambda$ doesn't represent a zero mode anymore, but a quasi-zero mode, since it parametrizes a family of local minima of the action when the constraint is imposed. Therefore, we still have the freedom of choosing the value that better fits the true Skyrmion energy. Also, even for a given, specific constraint, its value will not be unique, but will depend on the value of the other free parameter, namely, the gauge field mass $m$, which coincides with the (dimensionless) pion mass parameter of the Skyrme field. Therefore, there's a one parameter family of Skyrmion solutions, each with different values of $m$, and the best value of $\lambda$ for each one can be understood as a function $\lambda(m)$.

At the end of the day, constrained instantons also generate a seven dimensional Skyrmion moduli space, as fixing the Higgs vev at infinity does not break translation invariance, nor global gauge symmetry of the gauge field solution.

\begin{figure}[ht!]
    \centering
    \includegraphics[scale=0.6]{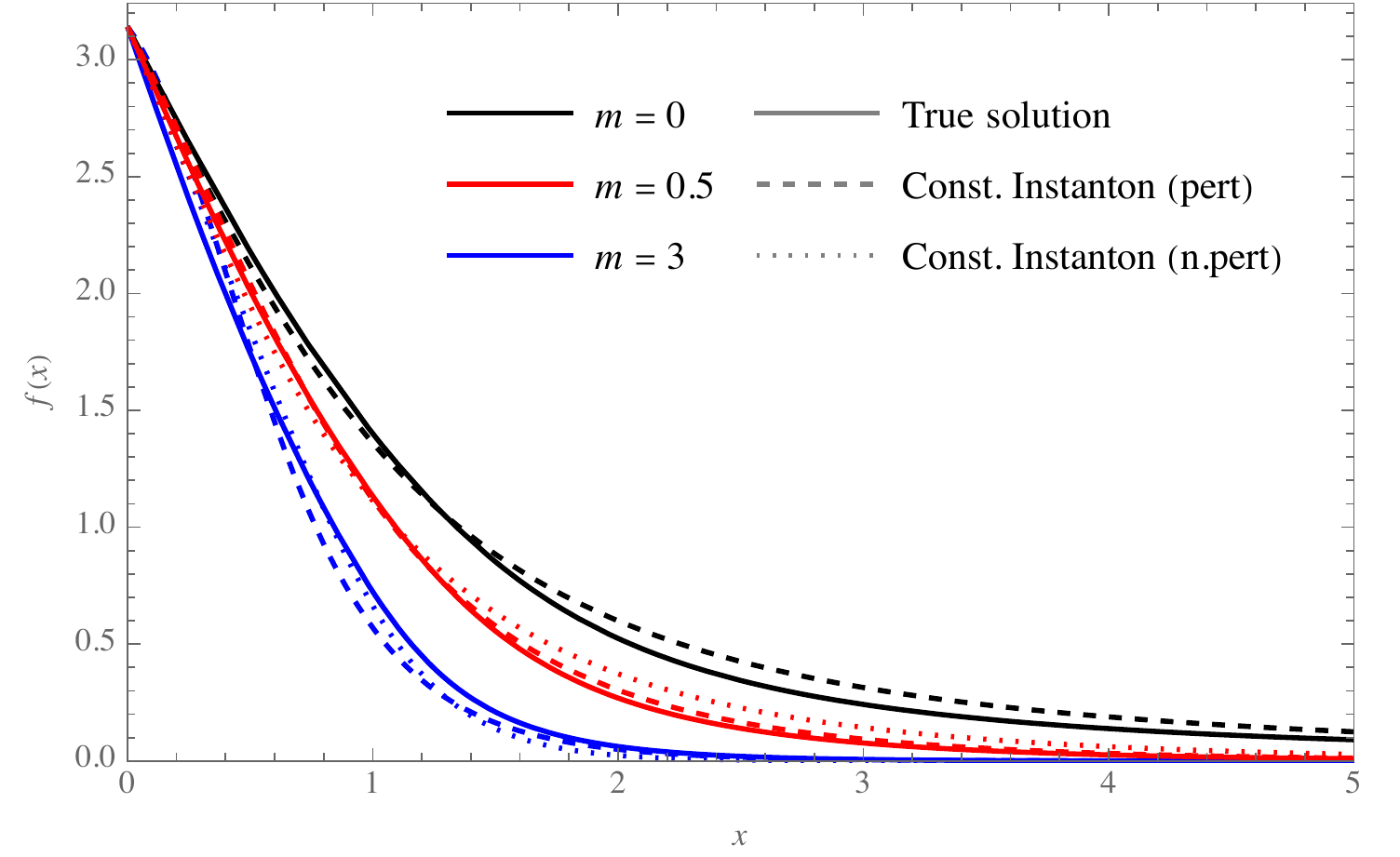}
    \caption{Solution of the Skyrme field profile for $m\in\{0,0.5,3\}$, and their corresponding best approximation using instantons (massless) and constrained instantons (massive) }
    \label{fig:constrained_instantons}
\end{figure}

As we can see from \cref{fig:constrained_instantons}, the similarity between the true solution in the massive pion case and the constrained instanton approximation is astounding, and even better than the regular instanton approximation in the massless case. Indeed, the difference in energies between both cases is smaller than $1\%$, for instance, $E_{m=0.5}=1.433\times 12\pi^2$ and $E^{\rm c.inst}_{m=0.5}=1.438\times 12\pi^2$. In this case, we have $\lambda^2(m=0.5)=1.47$. In \cref{tab:my_tabel} we show the energies for each case in units of $12\pi$. 
\begin{table}[h]
    \centering
    \begin{tabular}{c|c|c|c}
        $m$ &0&0.5&3  \\
        \hline
        $E_{\rm real}$ &$1.232$&$1.433$&$2.170$ \\
         $E^{\rm pert}_{\rm inst}$ & $1.243$ & $1.438$ & $2.250$ \\
                 $E^{\rm n.pert}_{\rm inst}$ & $1.243$  & $1.435$ & $2.224$ 
    \end{tabular}
    \caption{Energies of $B=1$ hedgehog obtained by numerically solving the Skyrme field equations ($E_{\rm real}$) and by the constrained instanton approximation, both using the perturbative configuration ($E^{\rm pert}_{\rm inst}$)
    and Wang's proposal ($E^{\rm n.pert}_{\rm inst}$).
    }
    \label{tab:my_tabel}
\end{table}

Now an interesting question is whether we can say anything about higher baryon number skyrmions. In the pure Yang-Mills case, the self duality of instantons allows for the construction of the most general $k$ instanton configuration via the powerful method of ADHM local data \cite{ADHM}. In \cite{Cork:2021uov}, this method was used to describe multi-Skyrmion moduli spaces in terms of the ADHM data of the corresponding instanton approximation.

In our case, however, self duality is broken, and the problem of finding expressions for $k>1$ constrained instantons, considerably harder. We could try to construct a $k$ constrained instanton as a superposition of $k$ individual 1-instantons, with arbitrary widths and positions, using a generalization of the 't Hooft ansatz to our case, i.e. taking \begin{equation}
    \alpha(x)=1+\sum\limits_{i=1}^k\frac{\lambda_i^2m}{|x-X_i|}K_1(m|x-X_i|),
\end{equation} 
where $\lambda_i$, $X_i$ correspond to the sizes and euclidean positions of each constrained instanton.
Unfortunately, this ansatz does not reproduce the full picture, as it does not allow for different orientations of the individual instantons. A modification of the 't Hooft ansatz that includes the relative orientation between instantons and does not modify the topological degree was proposed in \cite{Park:2002ie}. It's based on the introduction of $3k$ additional parameters, namely, the Euler angles of a $SU(2)$ rotation matrix $R_n$. Then, the fourth component of the gauge field is given by

\begin{equation}
        A_4(\mathbf{x},x_4)=-i\frac{1}{\alpha}\sum\limits_{i=1}^kR_i\tau_ a\partial_a\qty[\frac{\lambda_i^2m}{|x-X_i|}K_1(m|x-X_i|)]R^\dagger _i,
        \label{FinalA}
\end{equation}
which yields the correct number of collective coordinates on each topological sector. To what extent the gauge field configuration given in \cref{FinalA} does actually reproduce the moduli space of multi-Skyrmions in the massive pion case after calculating its holonomy along $x_4$ is out of the scope of this paper. Nevertheless, it is an interesting calculation that may allow us to study the effect of a nonzero pion mass in the Skyrmion-Skyrmion interactions. Such configuration, in the $k\rightarrow\infty$ limit could be used as well to approximate the (half-)Skyrmion crystals, since it is the constrained-instanton version of that proposed in \cite{Park:2002ie}. Indeed, a nonzero pion mass provides a natural cutoff for the individual instantons, which decay exponentially, and the holonomy can be calculated directly without needing to impose an artificial cutoff on instanton tails.

\section{Conclusions and outlook}
We have presented an approximation of Skyrmions in a model with massive pions using the holonomy of a constrained instanton of a Yang-Mills-Higgs theory in 4d Euclidean space. Constrained instantons are not even true minima of the Yang-Mills-Higgs action (due to the presence of a destabilizing scaling mode), and hence they are not unique. Nevertheless, they preserve their topological structure, so may still be useful to generate nontrivial Skyrmions via the Atiyah-Manton construction. Indeed, we proposed two of these configurations, namely, the linearized solutions of the equations of motion in their asymptotic form and the constrained instanton found in \cite{Wang:1994rz}, which are both easy to write and still allow to describe the correct asymptotic exponential behavior of the corresponding Skyrmion configuration. 

Therefore, we have shown that the relation between Euclidean Yang-Mills theory and the Skyrme model can be extended to account for massive pions by adding a scalar field with a Higgs potential to the Yang-Mills action. An interesting open question is whether this fact is a mere mathematical curiosity, or, on the contrary, it can be deduced from a holography-based argument, in the spirit of Sutcliffe's simple model. If this were the case, our finding may lead to new ideas on how to construct more realistic, bottom-up holographic duals for QCD that are able to describe explicit chiral symmetry breaking in the boundary theory. 

Obviously, having a short, semi-analytical approximation to the $k=1$ Skyrmion with massive pions can also be useful to study some dynamical aspects of the model without the need to rely in numerical methods to solve the corresponding Euler Lagrange equation. We would like to remark that a different method of generating Skyrmions with massive pions via computing the holonomy of instantons along circles in $\mathbb{R}^4$ was proposed in \cite{ATIYAH2005106}. The field configurations obtained this way correspond to Skyrmions with massless pions in hyperbolic space, which are shown to be a good approximation to Skyrmions in flat space with a nonzero pion mass, when the curvature parameter is related to the pion mass in a certain way. This idea results also in a very good approximation to the real solutions, but a drawback of this method is that the moduli space structure is not preserved, since the position of the skyrmion is related to a fixed point in the base manifold and it is not a free parameter (as there is not translation symmetry in hyperbolic space).
However, the instanton approximation is most useful for studying low energy dynamical processes involving two or more Skyrmions, since it allows to (approximately) describe the  corresponding moduli space. In principle, the extension of our approach to describe a multi-skyrmion moduli space is possible, although not straightforward. A first step would be to find constrained multi-instanton solutions, and study the corresponding moduli space. In this paper, we have proposed a simple configuration that is based on 't Hooft multi-instanton configuration, which is just a superposition of oriented constrained instantons. We believe this is a promising path towards a more realistic treatment of nuclear dynamics using the Skyrme model in the instanton approximation, since the effects of a nonzero pion mass should play a nontrivial role in Skyrmion-Skyrmion interactions.

\section*{Acknowledgements}
I would like to thank the organizers of the conferences SIG X in Krakow and Geometric Models of Nuclear Matter in Kent, for providing a friendly environment that allows the exchange of fresh ideas in the soliton community. I'd like to thank also J. Evslin for introducing to me the concept of constrained instantons, and to A.Wereszczynsky and M. Huidobro for reading the manuscript and providing useful comments.
This work has been done thanks to the funding of my predoctoral research activity from the Spanish Ministry of Science, Innovation and Universities, and the European Social Fund. This work has also received financial support from Xunta de Galicia (Centro singular de investigación de Galicia accreditation 2019-2022), by European Union ERDF, and by the “María de Maeztu” Units of Excellence program MDM-2016-0692 and the Spanish Research State Agency.

\bibliography{Bibliography.bib}

\end{document}